\documentclass[a4paper, 12pt]{article}

\usepackage{fancyhdr}
\usepackage{amsmath}
\usepackage{graphicx}
\usepackage{url}
\usepackage{amsbsy,amssymb} 
\usepackage{amsmath}
\usepackage{abstract}
\usepackage{hyperref}
\usepackage{color}
\usepackage{authblk}

\usepackage{fancyvrb}

\usepackage{rotating}

\usepackage[caption = false]{subfig}

\usepackage{lscape}

\title{A framework for on-line calibration of LINAC devices}
\author[1,2]{R. A. Kycia}
\author[1]{Z. Tabor}
\author[1]{A. Woszczyna}
\author[3]{D. Kabat}
\author[3]{M. Tulik}
\author[1]{Z. Lata\l a}
\affil[1]{Cracow University of Technology, Faculty of Physics, Mathematics and Computer Science, PL-31155, Krak\'ow, Poland}
\affil[2]{Department of Mathematics and Statistics, Masaryk University, Brno, Czechia}
\affil[3]{Centre of Oncology, Maria Sklodowska-Curie Memorial Institute, Krak\'ow, Poland}

\date{}
\begin{document}
\maketitle
\begin{abstract}
\noindent
General description of an on-line procedure of calibration for IGRT (Image Guided Radiotherapy) is given. The algorithm allows to improve targeting cancer by estimating its position in space and suggests appropriate correction of the position of the patient. The description is given in the Geometric Algebra language which significantly simplifies calculations and clarifies presentation.
\end{abstract}


\section{ Introduction }
Image Guided Radiotherapy (IGRT) is an image-based technique used to increase accuracy of ionizing radiation dose delivery during applications of fractions of a radiotherapy plan. In the most general terms, IGRT techniques can be divided into two classes: the first one includes procedures for correcting interfraction patient movements and the other one includes algorithms for correction of intrafraction movements, that is, the movements made during an actual therapy session. Both procedures use on-board imaging systems which are nowadays standard equipment of linear medical accelerators. The procedures in the former class usually involve acquisition of a computed tomography (CT) image of a patient just prior to dose delivery while the patient is already lying down on a treatment table. The CT image is compared to an image recorded during radiotherapy planning. The comparison returns patient positioning correction which must be made with a treatment table to align patient body organs in best accordance with their positions recorded during therapy planning. This class of IGRT corrections aims thus at decreasing patient positioning errors between successive therapy session. However, a patient can move also during a single therapy session (e.g. during breathing) and dose delivery procedures should account also for these movements through on-line calibration of the geometry of dose-delivery devices. Currently there is however no good solution for such on-line calibration procedures. The solutions which use calibration or patient positioning correction prior to the therapy and then assume that they are preserved during therapy \cite{Part1}, \cite{Part2GA} are good on average. Here we however present a procedure that increases the accuracy on-line, helps to target a cancer during dose delivery and is more computationally robust than usual techniques applied today. 

In particular, this paper presents a simple procedure for on-line calibration that ensures that the beam centre passes through a prescribed point in space such that its projection is visible by an imaging system. It employs usual setup during radiotherapy with additional non-invasive markers that are located close to the patient's body.

It seems that in the ease of derivations the most suitable language for geometrical considerations in (however not restricted to) three dimensional space is the Geometric Algebra (GA) developed by Clifford, Hilbert and Grassmann among others, see \cite{VectorWars}, and currently resurrected by David Hastenes \cite{Hastens_DesignOfAlgebra} and coworkers. It is a graded algebra over the real field. For a lengthy exposition see, e.g.,  \cite{Hastens_DesignOfAlgebra} or \cite{GeomALgForPhysicists} and applications to the projective geometry can be found in \cite{Hastens_Projective}. One may also consult Section 2 of \cite{Part2GA} for short exposition aimed at usefulness in cancer therapy. Short review will be also given in Section \ref{Section:Overview_of_the_Geometric_Algebra} below. Then we merge these ideas with the field of cancer therapy proposing on-line calibration procedure.

Th paper is organized as follows: the next section consists of short summary of useful formulas from the Geometric Algebra. Then the following section consists of three subsections that describe details of the proposed algorithm starting form preparation procedure that determines position of imaging and therapeutic subsystems. The algorithm assumes that we select only a point in a cancer that should be passed by an ionizing radiation beam central line. In the last section some consideration on the augmentation of the algorithm that take into account the tumour shape are provided.

\section{Overview of the Geometric Algebra}
\label{Section:Overview_of_the_Geometric_Algebra}
We start from reminding some standard results from GA, for elaborated discussion see, e.g., \cite{GeomALgForPhysicists}.

In the framework of geometric algebra for vectors $a$ and $b$ the geometric product is defined as
\begin{equation}
 ab = a\cdot b + a \wedge b,
 \label{Eq:GeometricProduct}
\end{equation}
where $a\cdot b$ is usual scalar product and $a \wedge b = -b \wedge a$ is the antisymmetric product - the wedge product - that results in a bivector of grade $2$, i.e., the plane spanned by $a$ and $b$ and oriented by the move from the first to the second component of the product. Higher graded components are obtained when successively perform wedge-multiplication and they also have simple geometric meaning.

\begin{figure}
\centering
 \includegraphics[width=0.5\textwidth]{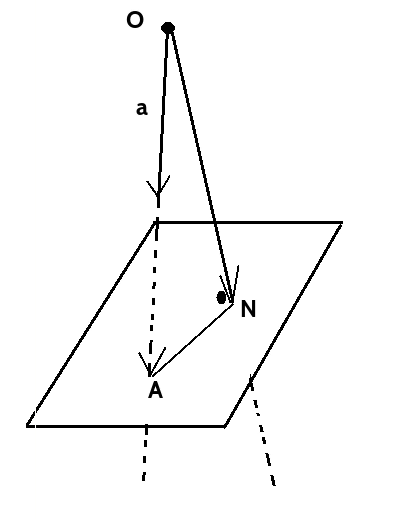}
 \caption{Projection of the point given by the vector $a$ onto the plane.}
  \label{Fig.ProjectivePlane}
\end{figure}

The most essential formula in cancer therapy is that which allows one to obtain a point $A$ by projecting a radiation source $O$ along a vector $a$ onto a plane $N=Ln$ - where a vector $N$ connecting the radiation source with the projection plane is perpendicular to the plane, a vector $n$ is unit: $n^{2}=1$ and therefore $L$ is the distance between the source and the imaging plane. These concepts are illustrated in Fig. \ref{Fig.ProjectivePlane}.
The formula for the projection is \cite{GeomALgForPhysicists, Part2GA}
\begin{equation}
 A-N = \frac{a \wedge N}{a \cdot N} N,
 \label{Eq:projectiveSplit}
\end{equation}
which can be further simplified, using (\ref{Eq:GeometricProduct}), to the well-known form
\begin{equation}
 A-N = \frac{aN^{2}-N(a \cdot N)}{a \cdot N}.
\end{equation}

In the Appendix of \cite{Part2GA} application of Mathematica CAS (Computer Algebra System) and the 'Cartan' package \cite{GAMathematica} is provided in order to perform efficient symbolic manipulations in Geometric Algebra. The Reader may consult it to avoid lengthy, however straightforward computations.

In the next section the description of the on-line calibration algorithm will be carefully outlined.

\section{Overview of the algorithm}
\begin{figure}
\centering
 \includegraphics[width=0.6\textwidth]{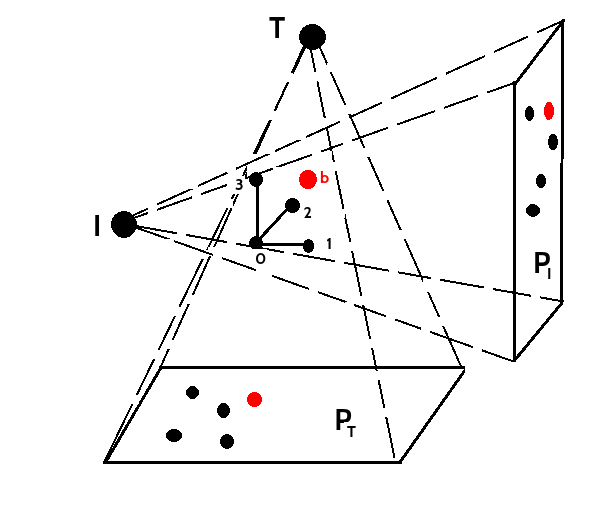}
 \caption{IGRT system. $I$ - an imaging system source; $P_{I}$ - an imaging plane for the imaging system; $T$ - a therapy source; $P_{T}$ - an imaging plane for the therapy system; $0,\ldots,3$ - the global reference frame - metal balls in positions $0$, $\{e_{i}\}_{i=1}^{3}$; $b$ - the position of an additional ball;}
  \label{Fig.IGRTSystem1}
\end{figure}

Fig. \ref{Fig.IGRTSystem1} presents an outline of a medical linear accelerator. Besides a Therapeutic system (T) it also consists of an X-ray scanning system (on-board imaging system)- called Imaging system (I) in the following which rotates around the same rotation axis as the Therapeutic system. (I) is rotated by 90 degrees related to (T). The long axis of a patient table is aligned with the axis of rotation of both subsystems. As the information from Image subsystem serves as an input for correction of Therapeutic system, therefore, it is important to impose strict constraints between (I) and (T) systems. The systems (I) and (T) (of different ionizing radiation energy) rotate around the common axis as a single part. The rotation trajectory usually deviates from ideal circular trajectory due to mechanical deformations of the supporting mechanisms due to enormous weights of accelerator parts. The angle of rotation is set through an operator console and it will be called $\alpha$. 

\begin{figure}
\centering
 \includegraphics[width=0.7\textwidth]{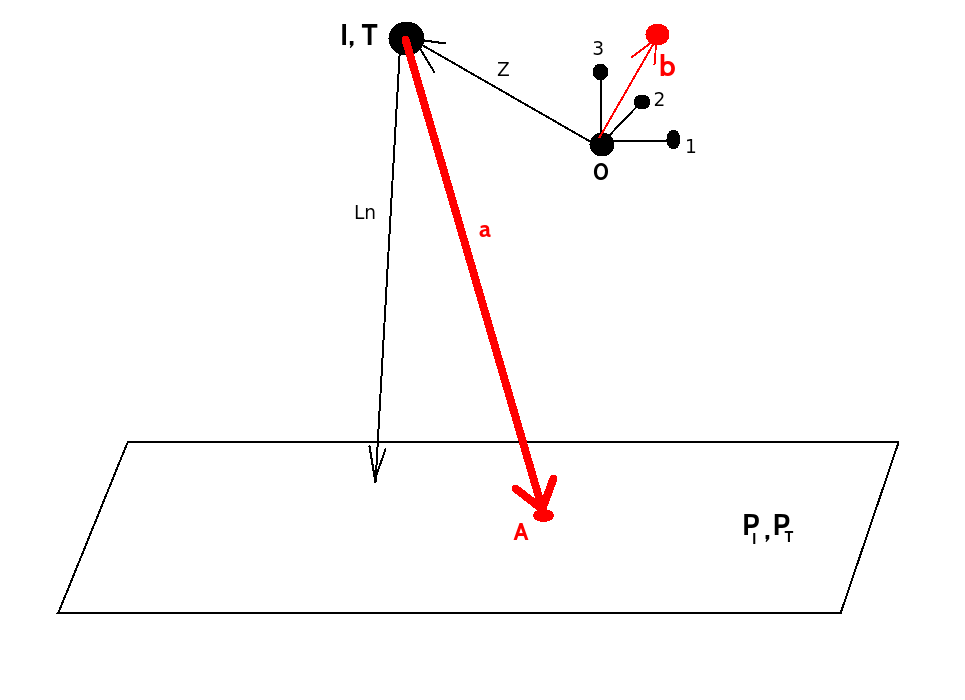}
 \caption{(I) or (T) subsystem. $I,T$ - an imaging/therapy system source; $P_{I},P_{T}$ - an imaging plane for the imaging/therapy system; $0,\ldots,3$ - the global reference frame - metal balls in positions $0$, $\{e_{i}\}_{i=1}^{3}$; $b$ - an additional ball position; $a$ - a vector of the centre of the ionizing radiation beam; $A$ - the projection of $a$ onto the imaging plane;}
  \label{Fig.OneSubsystem}
\end{figure}

To conduct on-line calibration a phantom is used consisting of metal balls located at $0$, $\{e_{i}\}_{i=1}^{3}$ and $b$. It is affixed to a treatment table and provides a global coordinate system.
The geometry of one of the subsystem is presented in Fig. \ref{Fig.OneSubsystem}.

The proposed algorithm for calibration consists of two steps:
\begin{enumerate}
 \item {Calibration of Imaging (I) and Therapeutic (T) (sub)systems and determination of their mutual location for a given setting of the rotation angle $\alpha$. This step is performed before the treatment.}
 \item {Application or radiotherapy using Imaging system and corrections of the patient position resulting from the proposed algorithm. This procedure is performed during the treatment for each angle $\alpha$ of the Gantry and Imaging system resulting from the therapy plan}
\end{enumerate}
The proposed algorithm is based  on the ideas presented in \cite{Part2GA} which are significantly extended here.

In the next subsections we present the algorithm in more detail.

\subsection{Calibration of (I) and (T) subsystems}
The first part of the algorithm is to calculate normal $n$ to an imaging plane, source to detector distance $L$, source position $Z$ and beam central vector $a$ for (I) and (T) subsystems independently. This procedure has to be performed before patient's treatment. It is assumed that the relations between these geometrical characteristics of (T) and (I) subsystems remain unchanged during the therapy. Otherwise, the notion of 'calibration' does not make sense.

The calibration method relays on introducing the new 'coordinate' system within a projective plane (either a detectors plane of (I) or (T) subsystem). This 'coordinates system' is given by the projection of the reference $3D$ coordinate system and selected special points (e.g. origin). It is quite similar in principle to introducing barycentric coordinate system and different than, well known, vector base decomposition. Therefore coordinates are the distances from fixed points given by mentioned special points projections. We call these distances 'projective coordinates'.

\begin{figure}
\centering
 \includegraphics[width=0.6\textwidth]{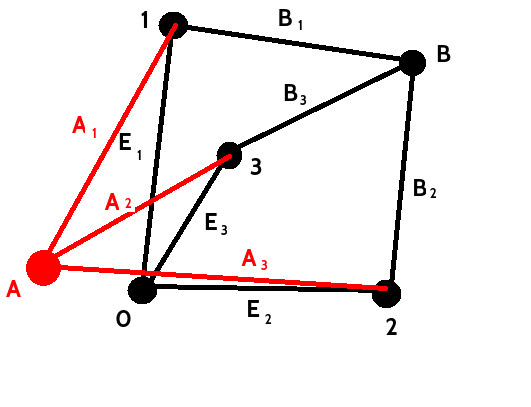}
 \caption{Measurements in calibration procedure.}
  \label{Fig.CalibrationMeasurements}
\end{figure}

The results of projection of the system from Fig. \ref{Fig.OneSubsystem} is presented in Fig. \ref{Fig.CalibrationMeasurements}. In order to derive $n$, $L$, $Z$, and $a$ one has to measure at least the distances $E_{i}$ and $A_{i}$ presented in Fig. \ref{Fig.CalibrationMeasurements} for (I) and (T) systems. The number of measurements must be big enough to invert the projection formulas listed below.

The projective coordinates follow from (\ref{Eq:projectiveSplit}), see also \cite{Part2GA}. For $e_{i}$ vectors the lengths of the projections are
\begin{equation}
 |E_{i}| = \left|\frac{(-Z+e_{i})\wedge n}{(-Z+e_{i})\cdot n} Ln - E_{0} \right|, \quad i \in \{1,2,3\},
 \label{Eq.Ei}
\end{equation}
where
\begin{equation}
 E_{0}=\frac{Z \wedge n }{Z \cdot n}Ln.
\end{equation}
The distance between a ball $b$ projection and the $e_{i}$ projections is given by
\begin{equation}
 |B_{i}| = \left|\frac{(-Z+b)\wedge n}{(-Z+b)\cdot n} Ln -  \frac{(-Z+e_{i})\wedge n}{(-Z+e_{i})\cdot n} Ln \right|,
 \label{Eq.Bi}
\end{equation}
Finally, the central axis of a beam intersects the projective plane in a point $A$ given by the distances
\begin{equation}
 |A_{i}|= \left|A- \frac{(-Z+e_{i})\wedge n}{(-Z+e_{i})\cdot n} Ln \right|, \quad i\in\{1,2,3\},
 \label{Eq.Ai}
\end{equation}
where
\begin{equation}
 A=\frac{a\wedge n}{a\cdot n} Ln.
\end{equation}
A comment on $A$ now is in order. It is some (arbitrarily chosen) point of the beam on the projective plane which we want to point exactly at the centre of a cancer as in \cite{Part2GA}. If the beam is non-uniform it can be naturally selected as the point in the maximum of intensity, otherwise it can be selected based on some geometrical characteristic of the beam, e.g., its geometrical centre.

The equations (\ref{Eq.Ei}), (\ref{Eq.Bi}) and (\ref{Eq.Ai}) allows one to derive $L_{\alpha}^{k}$, $n_{\alpha}^{k}$, $Z_{\alpha}^{k}$ and $a_{\alpha}^{k}$ for a given $\alpha$ value and  for $k\in\{(T),(I)\}$ for therapeutic and imaging system. It requires (usually numerical, as in \cite{Part1}) inversion and can be called 'unprojection'. This procedure can be schematically presented as
\begin{equation}
 (L_{\alpha}^{k}, n_{\alpha}^{k}, Z_{\alpha}^{k},  a_{\alpha}^{k}) = U(\{E_{i}^{k}\}_{i=1}^{3}, \{B_{i}^{k}\}_{i=1}^{3}, \{A_{i}^{k}\}_{i=1}^{3}),
\end{equation}
for $k\in\{(T),(I)\}$ subsytems, and where $U$ is the 'unprojection' operation. The numbers of variables on the left and the right site are the same, i.e., $9$ variables, however, in some numerical procedures of inverting of (nonlinear) projection equations more measurements (i.e. adding additional points $b$ in the phantom), may minimize the error \cite{Part1}.

The values of $L_{\alpha}^{k}$, $n_{\alpha}^{k}$, $Z_{\alpha}^{k}$ and $a_{\alpha}^{k}$ for a given $\alpha$ should be tabulated as they will be used as device characteristics for therapeutic on-line calibration during the patient treatment as it is described in the next subsection.

\subsection{On-line calibration during therapy}
During therapy the Imaging system (I) is used for alignment of the patient in order to correct for intrafraction patient movements and to minimise radiation dose delivery errors related to these movements. To account for the intrafraction movements information on localization of LINAC parts in space from the previous step, namely, $L_{\alpha}^{k}$, $n_{\alpha}^{k}$, $Z_{\alpha}^{k}$ and $a_{\alpha}^{k}$ for a given $\alpha$ value and  for $k\in\{(T),(I)\}$ has to be used.

In the second step of the algorithm the same phantom with balls as in the previous step is used. It is in the same position as during calibration and is fixed in space - it defines the global coordinate system. IGRT system is rotating around the patient table and the phantom. In addition, the patient table is allowed to move in order to apply alignment corrections as determined by the proposed algorithm. The primary goal of the corrections is to keep the beam central axis vector $a_{\alpha}^{T}$ directed towards a selected cancer point.

Hereafter, the system is fixed at $\alpha$ angle, and we will not use this parameter subscript if it is obvious. We will consider alignment for such fixed $\alpha$.

\begin{figure}
\centering
 \includegraphics[width=0.6\textwidth]{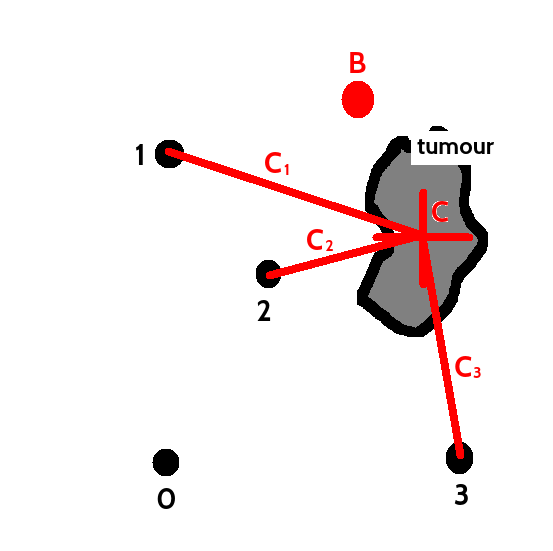}
 \caption{Imaging plane image during therapy. The radiotherapist marked the central point $C$ of the cancer, that should positioned in the central beam axis as specified by a vector $a_{\alpha}^{T}$. $0,1,2,3$ and $B$ are the markers of the phantom.}
  \label{Fig.ImagingnMeasurementsTherapy}
\end{figure}

The first stage is to use the Imaging system to mark the central cancer point $C$ on the image system plane as presented in Fig. \ref{Fig.ImagingnMeasurementsTherapy} and then measure distances from the projection of $C$ to at least three markers.
Given distances $|C_{i}|$ for $i\in\{1,2,3\}$ one can recover the vector $c^{I}$ connecting $I$ beam source with the cancer by reversing the projection equations
\begin{equation}
 |C_{i}|= \left|C- \frac{(-Z+e_{i})\wedge n}{(-Z+e_{i})\cdot n} Ln \right|, \quad i\in\{1,2,3\},
 \label{Eq.Ci}
\end{equation}
where
\begin{equation}
 C = \frac{c^{I} \wedge n^{I} }{ c^{I} \cdot n^{I}} L^{I}n^{I}.
\end{equation}

\begin{figure}
\centering
 \includegraphics[width=0.7\textwidth]{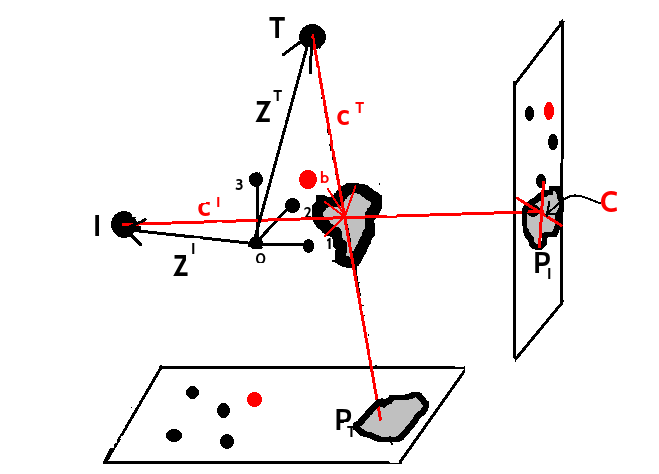}
 \caption{Cancer and phantom position in IGRT system.}
  \label{Fig.IGRTSystem2}
\end{figure}

Fig. \ref{Fig.IGRTSystem2} presents the whole system in space.
Knowing $c^{I}$ one can easily recover $c^{T} = c^{I}+Z^{I}-Z^{T}$. This vector now allows one to calculate the patient shift that transforms $c^{T}$ to a point on the line $a^{T}$ - that is the therapeutic beam centre line. The correction vector is the component of $c^{T}$ perpendicular to $a^{T}$, i.e.,

\begin{equation}
 \Delta C_{\alpha}=\frac{c^{T} \wedge a}{a^{2}}a,
 \label{Eq.RadiationSourceCorrection}
\end{equation}
which can be easily rewritten in a more familiar vector form
\begin{equation}
 \Delta C_{\alpha} = \frac{c^{T} a^{2}-(c^{T} \cdot a) a}{a^{2}}.
\end{equation}

This shift can be used for a patient alignment only for a fixed $\alpha$. Changing its value one has to repeat the procedure for shift calculation.

Up to now we considered the corrections assuming that there is a characteristic point in a cancer that has to be passed by the central axis of the beam. This assumption applies to cases of e.g. small cancer size, symmetric shape etc. In the next section we augment the proposed algorithm for the situations when a cancer tumour shape cannot be neglected.

\section{Shape of cancer tumour}

In this section we review standard techniques of forming the beam using multileaf collimator to fit the cancer shape. Then we describe a variant of above algorithm that simplifies calculations for such shape matching.

\subsection{Shape of a cancer - standard approach}

Medical accelerators are equipped with the first and the second stage collimator. The first collimator shapes the beam to the form of a pyramid using four movable massive metal blocks, so called jaws. The second stage collimator consists of multiple leaves which are simply metal plates of a few millimetre thickness. Each leaf can move independently on each other. There are two banks of leaves - the right bank and the left bank. Due to the construction of the multileaf collimator the  gap between the banks of leaves can be adjusted to any prescribed shape. 

During therapy planning a CT image of a patient is acquired and then a physician marks in 3D contours of a cancer and of organs at risk. Then, the physician with the help of a therapy planning system designs a therapy plan which consists of, among other issues, the leaves trajectories determined in such a way that the gap formed by the banks fits to the shape of a cancer for every angular orientation $\alpha$ of the therapeutic system included in the plan. The actual gap shape is calculated by projecting the 3D cancer shape on the projective plane for the given angular orientation $\alpha$ of (T).

\subsection{Shape of a cancer - GA approach}

To account for a cancer shape the precise shape of the radiation beam must be determined for every angular orientation $\alpha$ of (T). It requires in turn that the orientation of a 3D cancer object in space is restored from its projection onto the imaging plane $P_{I}$. If a cancer or organs next to it have some specific morphology that allows to distinguish some characteristic points inside them then one can use the base $1,2,3$ as for $|C_{i}|$ in (\ref{Eq.Ci}) to restore exact position of the cancer in space and if more points are used (at least 4 - the origin and three axes directions) then also tumour orientation can be determined. Then using geometric information on the liner accelerator geometric calibration from the algorithm presented above at fixed $\alpha$, exact beam shape can be automatically formed by projecting the cancer shape with already determined position and orientation onto the projective plane.

\begin{figure}
\centering
 \includegraphics[width=0.7\textwidth]{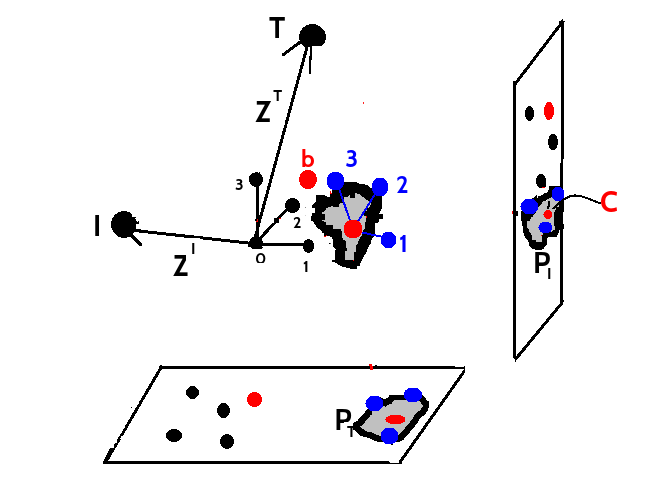}
 \caption{Cancer with some markers/characteristic points (blue dots). They are associated with some coordinate system and therefore determine orientation of the cancer in 3D space.}
  \label{Fig.ToumourSystem}
\end{figure}

Fig. \ref{Fig.ToumourSystem} presents the setup with markers/characteristic points. It can be compared with Fig. \ref{Fig.IGRTSystem2}.
The algorithm presented in the previous section can be used to determine the markers in the cancer $\{d_{i}\}_{i=1}^{3}$ (blue dots in Fig. \ref{Fig.ToumourSystem}). As from the previous step one knows the centre point of the cancer $c^{I}$ and $c^{T}$ the whole orientation of the cancer in 3D space is determined. Then adjustment by shifting the centre (\ref{Eq.RadiationSourceCorrection}) as well as change of the collimator leafs to account for the shape as projected onto the projective plane of (T) can then be easily performed by standard procedures.

\section{Summary}
In the paper the general on-line method of the beam alignment was presented which is based on some additional markers that carry a coordinate system on the projection plane of the detectors. 
The formulation was presented within the framework of Geometric Algebra, that provides efficient tools for calculation of projections. In addition, an alternative to current IGRT approach is presented. The method requires less computational cost as it bases only on a few characteristic points and not on the whole 3D image transformation as current IGRT methods. 
Due to their computational complexity current IGRT methods are used primarily for correcting interfraction patient movements \cite{IGRT_Corrections}. Very low computational cost of the proposed algorithm opens the doors for practical implementations of intrafraction IGRT.

\section*{Acknowledgments}
This research was supported by  the grant POIR.04.01.04-00-0014/16 of The National Centre of Research and Development.

\appendix




\end{document}